# Nonadditive statistical measure of complexity and values of the entropic index $q$

Sumiyoshi Abe[1], P. T. Landsberg[2], A. R. Plastino[3], and Takuya Yamano[4]

[1]*Institute of Physics, University of Tsukuba, Ibaraki 305-8571, Japan*

[2]*Faculty of Mathematical Studies, University of Southampton, Southampton SO17 1BJ, United Kingdom*

[3]*Physics Department, University of Pretoria, Pretoria 0002, South Africa*

[4]*Max-Planck-Institut für Physik Komplexer Systeme, Nöthnitzer Str. 38, D-01187 Dresden, Germany*

A two-parameter family of statistical measures of complexity are introduced based on the Tsallis-type nonadditive entropies. This provides a unified framework for the study of the recently proposed various measures of complexity as well as for the discussion of a whole new class of measures. As a special case, a generalization of the measure proposed by Landsberg and his co-workers based on the Tsallis entropy indexed by $q$ is discussed in detail and its behavior is illustrated using the logistic map. The value of the entropy index, $q$, with which the maximum of the measure of complexity is located at the edge of chaos, is calculated.

Given a probability distribution $p = \{p_i\}_{i=1,2,\ldots,W}$, with $W$ the number of accessible states at a certain scale, how can its configurational complexity be quantified? Both the completely ordered (CO) and completely random (CR) states are not complex. As physical examples, one can bear in mind the states of a perfect crystal and an ideal gas as the CO and CR states, respectively. These are rather simple states. Complex states exist between the CO and CR states. Therefore it seems natural to impose the condition that a statistical measure of complexity, $C$, if appropriately defined, should vanish at both the CO and CR states:

$$C[\mathrm{CO}] = C[\mathrm{CR}] = 0. \tag{1}$$

Obviously there are wide possibilities in specifying such a measure. This shows that the notion of complexity is still poorly determined. However, several proposals have been made in recent years for quantifying configurational complexity. Landsberg and his co-workers [1,2] have introduced a measure of disorder and then described complexity in terms of it. (See, also Refs. [3,4].) Using the model of an ideal nonequilibrium Fermi gas, they discussed the dependencies of disorder and complexity on temperature. López-Ruiz, Mancini and Calbet [5] have suggested instead the product of the Boltzmann-Shannon entropy of $p$ and the geometric distance between $p$ and the CR state. General properties of this measure have been studied in detail in Ref. [6]. Furthermore, in Refs. [7,8], the product of the Boltzmann-Shannon entropy of $p$ and the Kullback-Leibler entropy [9] of $p$ with respect to the CR state (chosen as a reference state) has been examined.

Thus, different statistical measures of complexity have been proposed and independently investigated in the literature. There are however two common features in the above approaches. One is that they all use the Boltzmann-Shannon entropy to realize the condition $C[\mathrm{CO}] = 0$. The other is that *they are nonadditive*, i.e., they do not possess additivity under replication of the system under consideration.

Now on the other hand, generalizations of the Boltzmann-Shannon entropy have been discussed in the area of nonextensive statistical mechanics. Among others, the entropy proposed by Tsallis [10] has been used to formulate generalized thermostatistics for analyzing observed phenomena and theories involving (multi)fractal structures, long-range interactions and long-time memories [11].

Taking into account nonadditive of the previously proposed statistical measures [1,2,5,7,8], it is of interest to examine possible roles of nonadditive entropies in quantifying configurational complexity. In this paper, we give a unified description of those proposed statistical measures of complexity using the Tsallis-type nonadditive entropies. We first define a two-parameter family of measures and show that all of them belong to this family. Then we consider a single-parameter subfamily that can be regarded as a generalization of the measure proposed by Landsberg and his co-workers, and study its properties employing the logistic map as a prototype example. We shall see how this measure behaves at the critical point of transition from regular to chaotic evolutions of the dynamical system. Special attention is paid to the dependence of the behavior of the measure on the parameter that characterizes the degree of nonadditivity.

We start our discussion by recalling the definition of the Tsallis entropy. For a probability distribution $p = \{p_i\}_{i=1,\,2,\ldots,\,W}$, it is given by

$$S_q[p] = \frac{k_B}{1-q}\left[\sum_{i=1}^{W}(p_i)^q - 1\right], \tag{2}$$

where $q$ is the positive entropic index and $k_B$ is the Boltzmann constant which is henceforth set equal to unity. Clearly, this quantity converges to the Boltzmann-Shannon entropy in the limit $q \to 1$:

$$S_q[p] \to S[p] = -\sum_{i=1}^{W} p_i \ln p_i. \tag{3}$$

$S_q$ possesses most of basic properties satisfied by the Boltzmann-Shannon entropy, such as the $H$-theorem and concavity. Additivity is however modified as follows:

$$S_q\left[p^{(1)}p^{(2)}\right] = S_q\left[p^{(1)}\right] + S_q\left[p^{(2)}\right] + (1-q)S_q\left[p^{(1)}\right]S_q\left[p^{(2)}\right]. \tag{4}$$

Thus, $S_q$ is nonadditive unless $q \to 1$. Like all entropies, the Tsallis entropy measures randomness. It vanishes for the CO state, in which $p_i = 1$ for a certain $i$ and the other probabilities are all zero. The CR state is the maximum Tsallis entropy state. It is the equiprobability state

$$p_i^{\mathrm{eq}} = 1/W \qquad (i = 1, 2, ..., W). \tag{5}$$

In this case, the Tsallis entropy is calculated to be

$$S_q\left[p^{\text{eq}}\right] \equiv S_q^{\text{eq}} = \frac{1}{1-q}\left(W^{1-q} - 1\right), \tag{6}$$

which tends to the celebrated Boltzmann formula $S^{\text{eq}} = \ln W$ in the limit $q \to 1$.

Next we quantify the difference between the configuration $p$ and the CR state. For this purpose, we propose to use the generalized Kullback-Leibler entropy [12-14]

$$K_q[p, p'] = \frac{1}{1-q} \sum_{i=1}^{W} (p_i)^q \left[(p_i)^{1-q} - (p_i')^{1-q}\right], \tag{7}$$

where $p'$ is the reference probability distribution. This quantity converges to the ordinary Kullback-Leibler entropy in the limit $q \to 1$:

$$K_q[p, p'] \to K[p, p'] = \sum_{i=1}^{W} p_i \ln \frac{p_i}{p_i'}. \tag{8}$$

$K_q[p, p']$ is positive semi-definite and is zero if and only if $p = p'$. Therefore it measures the difference between $p$ and $p'$. It can be shown that analogously to Eq. (4) $K_q$ satisfies the following pseudo-additivity relation:

$$K_q\left[p^{(1)} p^{(2)}, p'^{(1)} p'^{(2)}\right] = K_q\left[p^{(1)}, p'^{(1)}\right] + K_q\left[p^{(2)}, p'^{(2)}\right]$$
$$- (1-q) K_q\left[p^{(1)}, p'^{(1)}\right] K_q\left[p^{(2)}, p'^{(2)}\right]. \tag{9}$$

Generalizing the ideas proposed in Refs. [3,7,8], we now consider the following statistical measure of complexity:

$$C_{q,q'}[p] \equiv S_q[p] \cdot K_{q'}\left[p, p^{\mathrm{eq}}\right], \tag{10}$$

where the equiprobability distribution (the CR state) is chosen as the reference state. It is obvious that this quantity satisfies Eq. (1). At this stage, $q$ and $q'$ are taken to be different for the sake of generality. So, Eq. (10) defines a two-parameter family of measures. Since the second factor on the right-hand side of the above equation is

$$K_{q'}\left[p, p^{\mathrm{eq}}\right] = W^{q'-1}\left(S_{q'}^{\mathrm{eq}} - S_{q'}[p]\right) \tag{11}$$

with $S_{q'}^{\mathrm{eq}}$ given in Eq. (6), Eq. (10) can also be written as

$$C_{q,q'}[p] = W^{q'-1} S_q[p]\left(S_{q'}^{\mathrm{eq}} - S_{q'}[p]\right). \tag{12}$$

Let us consider some limiting cases of $C_{q,q'}$ and see how the measures proposed in Refs. [1,2,5,7,8] are reproduced. First of all, recall that the measure given in Refs. [7,8] is the product of the Boltzmann-Shannon entropy and the ordinary Kullback-Leibler entropy with the equiprobability reference state. Therefore, clearly it is obtained in the limits $q, q' \to 1$:

$$C_{q,q'}[p] \to C_{1,1}[p] = S[p]\left(S^{\mathrm{eq}} - S[p]\right), \tag{13}$$

where $S^{\mathrm{eq}} = \ln W$. In terms of Landsberg's disorder [1]

$$\Delta[p] = \frac{S[p]}{S^{\mathrm{eq}}}, \tag{14}$$

$C_{1,1}$ is written as

$$C_{1,1}[p] = \left(S^{\mathrm{eq}}\right)^2 \Delta[p]\left(1 - \Delta[p]\right). \tag{15}$$

Up to the constant prefactor $\left(S^{\mathrm{eq}}\right)^2$, this quantity is equivalent to the measure of complexity proposed in Ref. [2]

$$\Gamma[p] = \Delta[p]\left(1 - \Delta[p]\right). \tag{16}$$

Secondly, let us consider the limits $q \to 1$, $q' \to 2$. In this case, Eq. (11) becomes

$$\begin{aligned} K_{q'}\left[p, p^{\mathrm{eq}}\right] \to K_2\left[p, p^{\mathrm{eq}}\right] &= W\left[\sum_{i=1}^{W} \left(p_i\right)^2 - \frac{1}{W}\right] \\ &= W\sum_{i=1}^{W} \left(p_i - \frac{1}{W}\right)^2. \end{aligned} \tag{17}$$

The summation in this equation is the squared geometric distance between $p$ and $p^{\mathrm{eq}}$. Therefore, in these limits, we have

$$C_{q,q'}[p] \to C_{1,2}[p] = W\,S[p]\sum_{i=1}^{W}\left(p_i - \frac{1}{W}\right)^2 . \tag{18}$$

Up to the constant prefactor $W$, this is equivalent to the measure proposed by López-Ruiz, Mancini and Calbet in Ref. [5]. Thus we observe that all the proposed statistical measures of complexity mentioned here are unified in the present two-parameter family $\{C_{q,q'}\}$.

Moreover, it is of interest to consider a single-parameter subfamily, in which $q = q'$. In this case, Eq. (10) becomes

$$C_{q,q}[p] = W^{\,q-1}\left(S_q^{\mathrm{eq}}\right)^2 \Gamma_q[p], \tag{19}$$

where

$$\Gamma_q[p] = \Delta_q[p]\left(1 - \Delta_q[p]\right), \tag{20}$$

$$\Delta_q[p] = \frac{S_q[p]}{S_q^{\mathrm{eq}}}. \tag{21}$$

$\Delta_q$ and $\Gamma_q$ are the $q$-generalizations of Eqs. (14) and (16), respectively. Clearly, the maximum of $\Gamma_q$ is given by

$$\Gamma_q^{\max} = \Gamma_q\left(\Delta_q = 1/2\right) = \frac{1}{4}. \tag{22}$$

The maximization condition $\Delta_q = 1/2$, that is,

$$S_q[p] = \frac{1}{2} S_q^{\mathrm{eq}} \tag{23}$$

should fix the value of $q \equiv q*$.

This enables us finally to locate the maximum of complexity at the edge of chaos. In what follows, we study the behavior of $C_{q,q}$ employing the logistic map as a prototype example. We examine several values of $q$. We shall numerically determine the value of $q$, with which the maximum value, $C_{q,q}^{\max}$, is located at the point of transition from regular to chaotic evolutions. Note that $C_{q,q}^{\max}$ monotonically increases with respect to $W$.

So, let us consider the logistic map in the following representation [15]:

$$x_{n+1} = 1 - a\, x_n^2, \tag{24}$$

where $a \in [0, 2]$ is the control parameter and $x_n \in [-1, 1]$ ($n = 0, 1, 2, ...$). Following standard procedures in symbolic dynamics [15], we have translated each orbit $\left(x_1, x_2, x_3, ...\right)$ of the map into a binary sequence of 0's and 1's, according to the rule: 0 if $x \le 0$ and 1 if $x > 0$. As was done in Refs. [5,6], sequences of length 12 are regarded as the states of the system. That is, each string of 12 symbols is regarded as one possible state. The present measure of complexity has been evaluated on a probability distribution associated with this space of $2^{12}$ states. In order to find the concomitant probabilities, $p_i$ ($i = 1, 2, ..., 2^{12}$), we have iterated the map $12 \times 10^6$ times, thus generating $10^6$ (non-overlapping) sequences of length 12. The probability, $p_i$, associated with the $i$th state is, after proper normalization, proportional to $n_i$, which is the number of times of appearance of the *i*th state during the iteration of the map. Thus, the probability, $p_i$, measures how often that particular string of 12 symbols is "visited" during time evolution of the system. We have computed the value of $q$ at the critical point, at which the measure $C_{q,q}$ in Eq. (19) takes its maximum value; that is, the value of $q$, for which the quantity, $\Delta_q$, is equal to $1/2$. We have considered the critical point associated with the onset of chaos after the period doubling sequence. Such a critical value of the map parameter is given by [15]

$$a_c = 1.401155189.... \tag{25}$$

In Fig. 1, we present plot of $p_i$ at $a = a_c$. We have numerically calculated the associated *optimal* value of the parameter, $q = q^*$. It is found to be given by

$$q^* = 1.433601669.... \tag{26}$$

The present measure with this value of the index of nonadditivity can be thought of as an extreme improvement of the measures of complexity introduced in Refs. [3-5,7,8], since it takes its maximum value at the edge of chaos.

In Fig. 2, we present the plot of $C_{q^*,\,q^*}$. Its maximum at $a_c$ is clearly appreciated. (It is interesting to observe that $C_{q^*,\,q^*}$ takes its maximum also for the values of *a* different from $a_c$. We hope to analyze this point elsewhere.) For comparison, we also present the plot of $C_{1,1}$, which is essentially the measure given in Refs. [1,2]. Clearly, $C_{1,1}$ does not take its maximum at the edge of chaos.

In conclusion, we have proposed a unified description of configurational complexity based on a two-parameter family of statistical measures, which are essentially nonadditive. In particular, we have studied in detail a peculiar one-parameter sub-family of measures indexed by $q$, which is a generalization of the measure introduced in Refs. [1,2]. We have examined it for the logistic map as a prototype example and have determined the associated value of the index, $q$. Further applications of the present measure to other dynamical systems are welcome.

*Note added*. In their recent paper [16], Piasecki, Martin and Plastino have made use of the Tsallis entropy to quantify inhomogeneity and complexity of spatial patterns. Also, one of us (T. Y.) has recently examined the use of generalized entropies for measuring statistical complexity [17]. However, the scopes of these works are different from the present one.

**Acknowledgments**

S. A. was supported in part by the Grant-in-Aid for Scientific Research of the Japan Society for the Promotion of Science. T. Y. thanks partial support from the Research Fellowships of the Japan Society for the Promotion of Science for Young Scientists.

# Figure Captions

Fig. 1 The probability distribution with $2^{12}$ states of the logistic map at the critical point $a_c$. Iteration of $12 \times 10^6$ times is performed, generating $10^6$ sequences of length 12. The initial condition is taken to be $x_0 = 0.45$. All quantities are dimensionless.

Fig. 2 The plot of the measure $C_{q^*, q^*}$ with $q^*$, fixed by the condition in Eq. (23), calculated by using the distribution in Fig. 1. All quantities are dimensionless.

Fig. 3 The plot of $C_{1,1}$ with respect to *a*. All quantities are dimensionless.

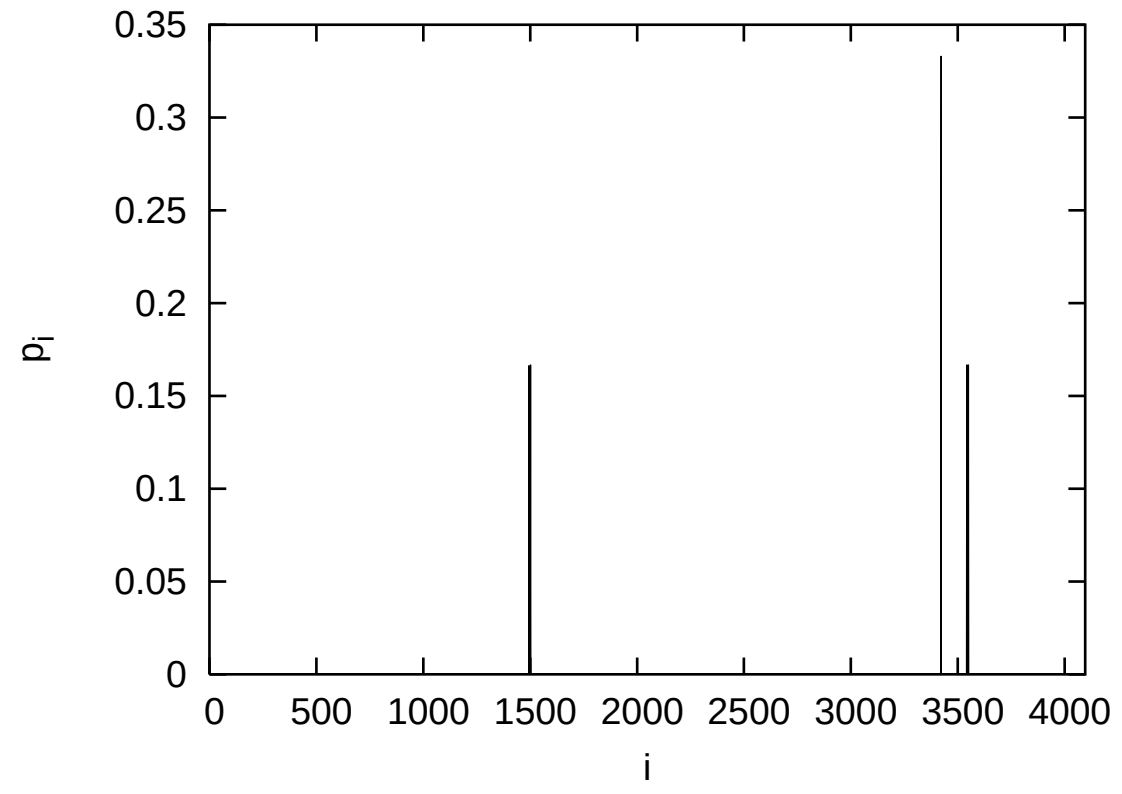

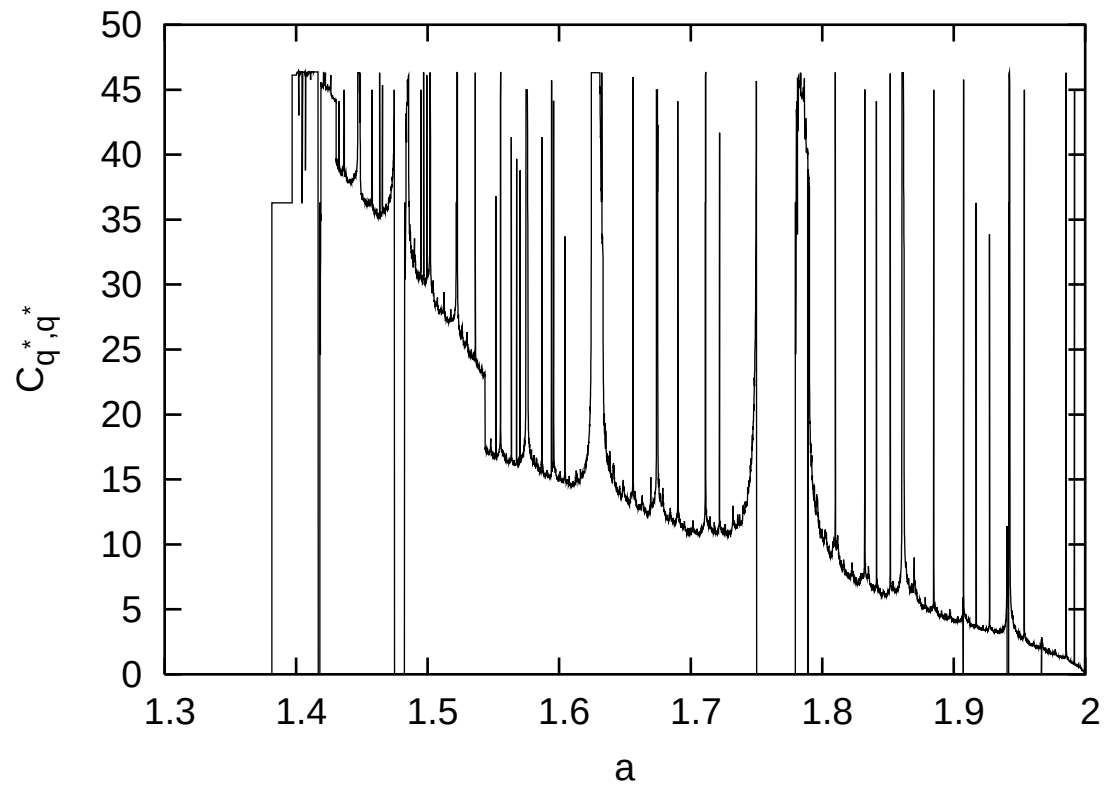

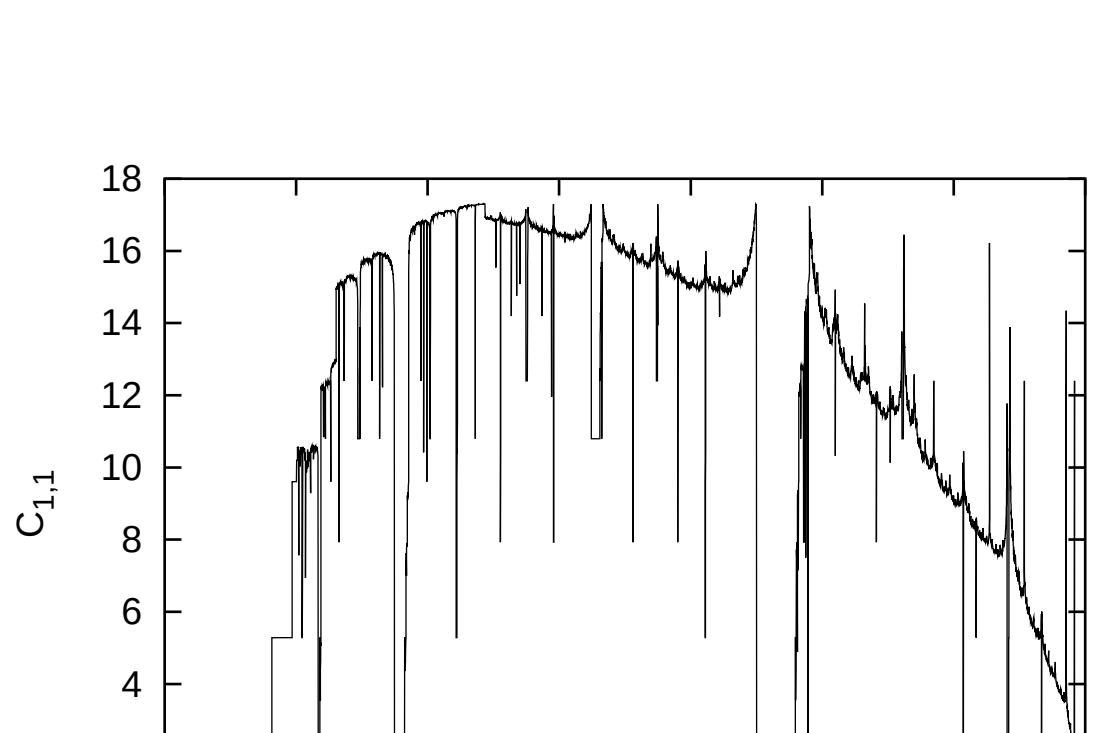